\documentclass[aps,twocolumn,groupedaddress,floatfix]{revtex4-1}
\usepackage{amssymb,amsmath}
\usepackage{graphicx}
\usepackage{subfigure}
\usepackage[english]{babel}
\usepackage{float}
\usepackage{color}

\usepackage{lipsum}
\usepackage{suffix}
\usepackage{mathtools}

\DeclarePairedDelimiterX\MeijerM[3]{\lparen}{\rparen}%
{\begin{smallmatrix}#1 \\ #2\end{smallmatrix}\delimsize\vert\,#3}

\newcommand\MeijerG[8][]{%
  G^{\,#2,#3}_{#4,#5}\MeijerM[#1]{#6}{#7}{#8}}

\WithSuffix\newcommand\MeijerG*[7]{%
  G^{\,#1,#2}_{#3,#4}\MeijerM*{#5}{#6}{#7}}
\begin{document}
\newcommand{\be}{\begin{equation}}
\newcommand{\ee}{\end{equation}}
\newcommand{\rojo}[1]{\textcolor{red}{#1}}
\newcommand{\azul}[1]{\textcolor{blue}{#1}}

\title{A two-dimensional disordered magnetic metamaterial}
\author{Cristian Mej\'{\i}a-Cort\'{e}s}
\affiliation{Programa de F\'{\i}sica, Facultad de Ciencias B\'{a}sicas, Universidad del Atl\'{a}ntico, Puerto Colombia 081007, Colombia}
\author{Mario I. Molina}
\affiliation{Departamento de F\'{\i}sica, Facultad de Ciencias, Universidad de Chile, Casilla 653, Santiago, Chile}

\date{\today }

\begin{abstract} 

We study the effect of a resonant frequency disorder on the eigenstates and the transport of magnetic energy in a two-dimensional (square) array of split-ring resonators (SRRs). In the absence of disorder, we find the dispersion relation of magneto-inductive waves and the mean square displacement (MSD) in closed form, showing that at long times the MSD is ballistic. When disorder is present, we consider two types: the usual Anderson distribution (uncorrelated monomers) and $2 \times 2$ units assigned at random to lattice sites (correlated tetramers). This is a direct extension to two dimensions of the one-dimensional random dimer model (RDM). For the uncorrelated case, we see saturation of the MSD for all disorder widths, while for the correlated case we find a disorder window, inside which the MSD does not saturate at long times, with an asymptotic sub-diffusive behaviour $MSD\sim t^{0.26}$. Outside this disorder window, the MSD shows the same kind of saturation as in the monomer case. We conjecture that the sub-diffusive behaviour is a remanent of a weak  resonant transmission of a 2D plane wave across a tetramer unit.

\end{abstract}

\maketitle

{\em Introduction}.
The current ability to tailor material properties at will has lead to a whole class of artificial materials, termed metamaterials, characterized by unprecedented thermal, optical, and transport properties that make them attractive candidates for current and future technologies. Among them, we have magnetic metamaterials (MMs) that consist of an array of metallic split-ring resonators (SRRs) coupled inductively~\cite{SRR2, SRR3, SRR1}. This type of system can, for instance, feature negative magnetic response in some frequency window, making them attractive for use as a constituent in negative refraction index materials~\cite{padilla,veselago,pendry,negative_refraction}. 
The usual theoretical treatment of such structures is an effective medium approximation where the composite is treated as a homogeneous and isotropic medium, characterized by effective macroscopic parameters.  Of course, this approach is valid, as long as the wavelength of the external electromagnetic field is much larger than the linear dimensions of the MM constituents.

\begin{figure}[t]
 \includegraphics[scale=0.17]{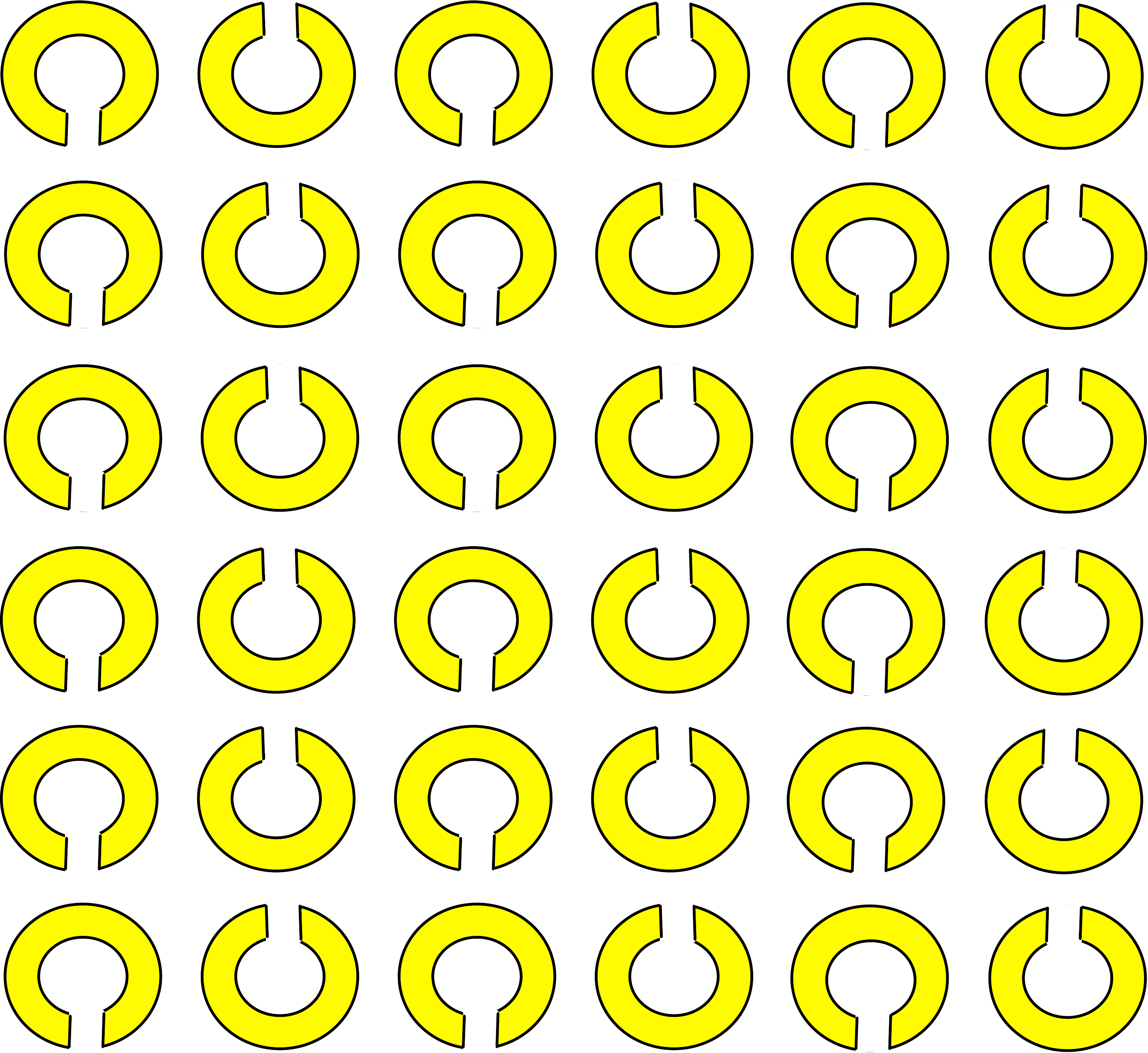}
  \caption{Two-dimensional split-ring resonator array.}
  \label{fig1}
\end{figure}

One of the simplest two-dimensional MM models consists of a periodic square array of split-ring resonators (SRRs) lying on a common plane, where each resonator consists of a small, conducting ring with a slit (Fig.~\ref{fig1}). Each SRR unit in the array is equivalent to a
resistor-inductor-capacitor (RLC) circuit featuring self-
inductance $L$, ohmic resistance $R$, and capacitance $C$
built across the slit~\cite{elefterio1,elefterio2}. If we assume a negligible resistance, 
each unit will possess a resonant frequency $\omega=1/\sqrt{L C}$.
Under this condition and in the absence of driving, the evolution equations for the charge $Q_{n}$ residing at the $nth$ ring are given by
\begin{eqnarray}
{d Q_{\bf n}\over{d t}}&=& I_{n}\\
L {d I_{\bf n}\over{d t}} + {Q_{\bf n}\over{C}}&=& -M\ \sum_{\bf m} {d I_{\bf m}\over{d t}}
\end{eqnarray}
where $M$ is the mutual inductance and the sum is restricted to nearest-neighbors of ${\bf n}$. These equations can be cast in dimensionless form as
\be
{d^2\over{d t^2}}\left( q_{\bf n} + \lambda\ \sum_{\bf m}q_{\bf m} \right) + \omega_{\bf n}^{2} q_{\bf n} = 0 \label{eq1}
\ee
where $q_{\bf n}$ denotes the dimensionless charge of the $nth$ ring, ${\bf n}=(n_{x}, n_{y})$,  
$\lambda$ is the coupling between neighboring rings that originates from the dipole-dipole interaction, and $\omega_{\bf n}^2$ is the (square of) resonant frequency of the $nth$ ring, normalized to a characteristic frequency of the system, 
$\langle \omega_{\bf n}^2\rangle=(1/N) \sum_{n} \omega_{\bf n}^2$. These equations can also be derived from
the Hamiltonian $H = \sum_{\bf n} H_{\bf n}$ where $H_{\bf n}$ is the Hamiltonian density
\be
H_{\bf n} = \frac{1}{2} \left(\dot{q}_{\bf n}^{2} + 
\lambda \dot{q}_{\bf n} \sum_{\bf m} \dot{q}_{\bf m} + \omega_{\bf n}^2 q^2_{\bf n} \right),\label{Hn}
\ee
where $\dot{q_{\bf n}}\equiv (d/d t) q_{\bf n}$. We assume that the magnetic component of any incident external electromagnetic wave is perpendicular to the SRRs' plane and that the electric field of the incoming wave is perpendicular to the electric field induced along the slits. This insures that only the magnetic component of the incoming wave creates an electromotive force on the rings, giving rise to an oscillating current in each SRR and to an oscillating voltage difference across the slits. Also, it is a good idea to reduce electric dipole-dipole effects coming from the strong electric fields at the slits by a judicious placing  of the SRRs in the common plane to keep the slit-to-slit distance as large as possible (Fig.~\ref{fig1}). However, large Ohmmic and radiative losses remain as the main drawback of the SRR array. A possible way to deal with this problem that has been considered is to endow the SRRs with external gains, such as tunnel (Esaki) diodes~\cite{losses1,losses2} to compensate for such losses.

The dimensionless stationary state equation is obtained from Eq.(\ref{eq1}) after
posing $q_{\bf n}(t) = q_{\bf n} \exp[i (\Omega t + \phi) ]$:
\be
-\Omega^2 \left( q_{\bf n} + \lambda \sum_{\bf m} q_{\bf m}\right) + \omega_{\bf n}^{2} q_{\bf n}=0.
\label{eq2}
\ee
The frequency $\omega_{n}^2$ can be changed by varying the capacitance of the ring, which is accomplished by altering the slit width or by inserting a dielectric in the slit. For a homogeneous array, $\omega_{n}^2=1$.

On the other hand, the effect of disorder on the stationary and transport properties of a discrete, periodic system, is an old topic, but its importance has not  waned throughout the years due to its fundamental importance in several fields. The most relevant result in this area is Anderson localization which asserts that the presence of disorder tends to inhibit the propagation of excitations. In fact, for 1D systems, all the eigenstates are localized and transport is completely inhibited~\cite{anderson1, anderson2, anderson3}. This was also proven to be true for two-dimensional systems, while in three-dimensions a mobility edge is formed.
Now, Anderson localization is based on the notion that the disorder is ``perfect'' or uncorrelated. However, it has been noted that in one-dimensional lattices with a correlated disorder, a degree of transport is still possible. This happens, for instance, in the random dimer model (RDM) for the usual tight-binding model. It consists of a binary lattice for the site energies where site energy is assigned at random to pairs of lattice sites. This leads to a mean square displacement of an initially localized excitation that grows asymptotically as $t^{3/2}$ at low disorder levels, instead of the saturation behavior predicted by Anderson's theory~\cite{correlated1,correlated2,correlated3}. 

An experimental demonstration of the RDM prediction has been made in an optical setting~\cite{szameit}. A straightforward extension of these ideas to random arrays of larger units (n-mers), has also been theoretically explored~\cite{nmers}. For a disordered, one-dimensional  SRR array, it was found that an uncorrelated disorder always leads to localization of magnetic energy at any disorder strength, with a transmission that decreases exponentially with the size of the system. For correlated disorder and small and medium disorder levels, however,  it becomes possible for a fraction of states to have resonant transmission, leading to a power-law decrease of the overall transmissivity with system size~\cite{arxiv}.

In this work, we examine the localization of the magnetic modes and the transport of magnetic excitations in a two-dimensional disordered square array of split-ring resonators, where the resonant frequencies $\omega_{\bf n}^2$ in Eq.(\ref{eq2}) are taken as random quantities. We will consider two cases: A completely uncorrelated  one where the $\omega_{\bf n}^2$ are assigned at random to individual units, or `monomers' and a correlated case, consisting of a straightforward 2D generalization of the well-known random dimer model used in one-dimensional systems: Site frequencies $\omega_{\bf n}^2$ assigned at random to $4$ nearby units, or `tetramers' (Fig.2)\cite{Naether}. 
\begin{figure}[t]
 \includegraphics[width = 1\columnwidth]{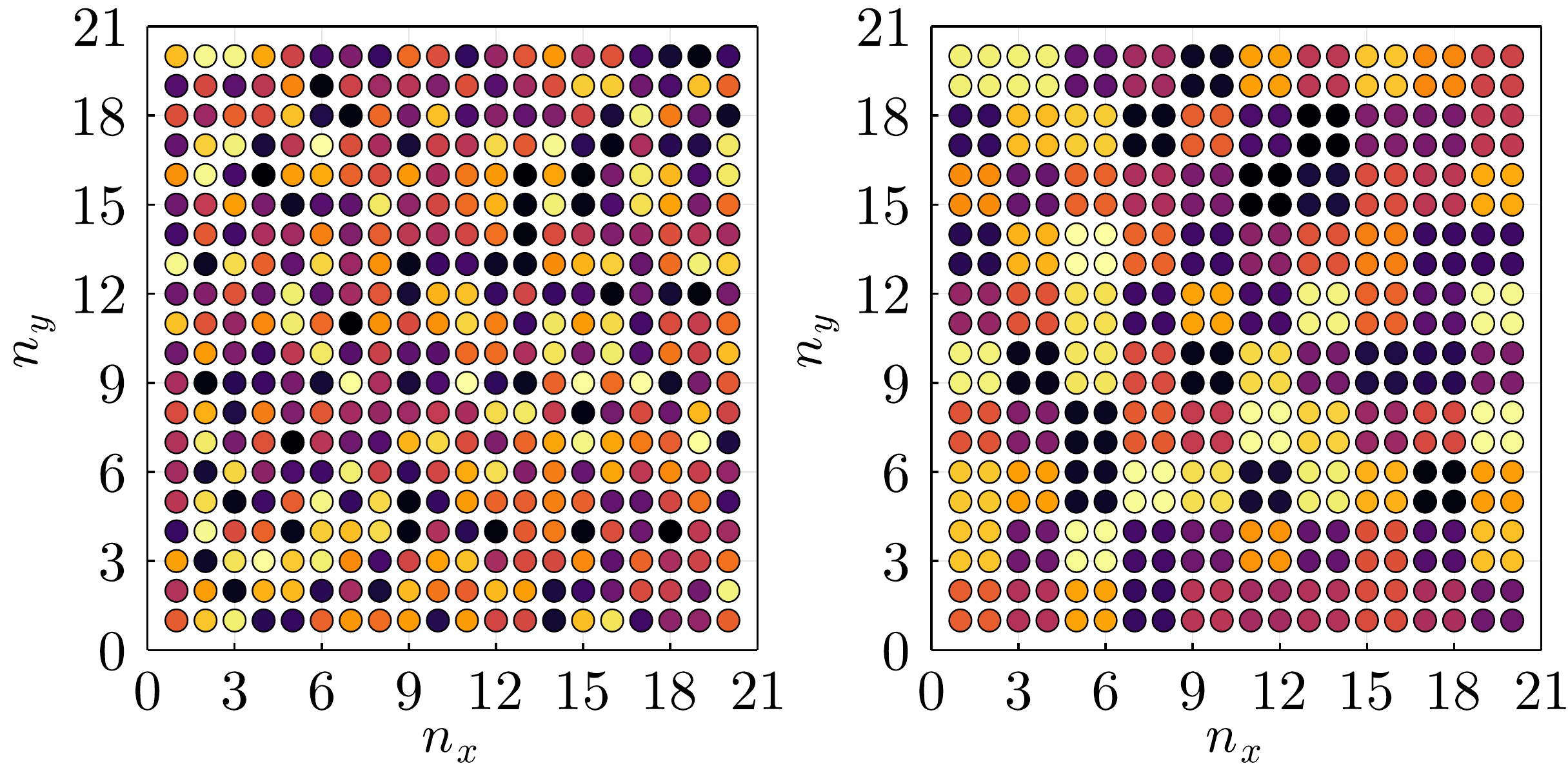}
\caption{Random realization of an uncorrelated (left) and a correlated (right) distribution of random resonant frequencies.}
  \label{fig2}
\end{figure}

A usual indicator of localization is given by the participation ratio (PR), that measures the extent of the electric charge 
distribution stored in the capacitors (or magnetic energy density stored in the inductors):
\be
PR = \left(\ \sum_{\bf n} |q_{\bf n}(t)|^2 \ \right)^2/\sum_{\bf n} 
|q_{\bf n}(t)|^4\label{PR}
\ee
For a completely localized excitation, $PR=1$, while for a complete delocalized state, $PR=N$.

To monitor the degree of mobility of a magnetic excitation we resort to the mean square displacement (MSD) of the charge, defined as
\be
\langle {\bf n}^2 \rangle = \sum_{\bf n} {\bf n}^2 |q_{\bf n}(t)|^2 / \sum_{\bf n} |q_{\bf n}(t)|^2\label{MSD}
\ee
Typically $\langle {\bf n}^2 \rangle\sim t^\alpha$ at large $t$, where $\alpha$ is known as the transport exponent. The types of motion are classified according to the value of $\alpha$: `localized' ($\alpha=0$), `sub-diffusive' ($0<\alpha<1$), `diffusive' ($\alpha=1$), `super-diffusive' ($1<\alpha<2$) and `ballistic' ($\alpha=2$).

{\em Homogeneous case}. Before embarking into the effects of disorder on the system, let us begin by examining the phenomenology in the absence of disorder, $\omega_{n}^2 = 1$, in order to have a proper comparison context. 

After posing $q_{\bf n}\sim \exp(i {\bf k}\cdot{\bf n})$ and solving for $\Omega^2$, we obtain the dispersion relation in $d$-dimensions as
\be
\Omega_{\bf k}^2 = {1\over{1 + \lambda\ \sum_{\bf m}\exp(i {\bf k}\cdot {\bf m})}}\label{om2}
\ee
where the sum is restricted to nearest neighbors. 

The time evolution of a completely localized initial charge $q_{\bf n}(0)=A\delta_{{\bf n},0}$, and no currents, $(d q_{\bf n}(0)/d t)=0$, is given by 
\begin{eqnarray}
q_{\bf n}(t) &=& (A/v) \int_{FBZ}  e^{i ({\bf k}\cdot {\bf n}-\Omega_{\bf k} t)} d{\bf k} \nonumber \\
&+& (A/v) \int_{FBZ} 
e^{i ({\bf k}\cdot {\bf n}+\Omega_{{\bf k}} t)} d{\bf k}
\end{eqnarray}
where $v$ is the volume of the first Brillouin zone (FBZ), and $\Omega_{\bf k}^2$ is given by Eq.(\ref{om2}). After replacing this form for $q_{\bf n}(t)$ into Eq.(\ref{MSD}), one obtains after some algebra, a closed form expression for $\langle {\bf n}^2 \rangle$:
\be 
\langle {\bf n}^2 \rangle = {  (1/v) \int_{FBZ} (\nabla_{\bf k} \Omega_{\bf k})^2 (1 - \cos(2\ \Omega_{\bf k}\ t))
\over{
1 + (1/v)\ \int_{FBZ} \cos(2\ \Omega_{\bf k}\ t)}
}\ t^2.\label{eq9}
\ee
At long times $\langle {\bf n}^2 \rangle$ approaches a ballistic behavior
\be
\langle {\bf n}^2 \rangle =\left[\ {1\over{v}} \int_{FBZ} (\nabla_{\bf k} \Omega_{\bf k})^2 d{\bf k}\ \right]\ t^2\hspace{1cm} (t\rightarrow \infty)
\ee
while at short times,
\be
\langle {\bf n}^2 \rangle = \left[\ {1\over{v}} \int_{FBZ} \Omega_{\bf k}^2 (\nabla_{\bf k} \Omega_{\bf k})^2 d{\bf k}\ \right] \ t^4 \hspace{1cm}(t\rightarrow 0).
\ee

For a square lattice ($d=2$),
\be
\Omega_{\bf k}^2 = {1\over{1 + 2 \lambda \left[\cos(k_{x})+\cos(k_{y})\right]}}\label{om22}
\ee
where, ${\bf k}=(k_{x},k_{y})$. We see that the system is capable of supporting magnetoinductive waves, if $|\lambda|<1/4$.
\begin{figure}[t]
 \includegraphics[width=1\columnwidth]{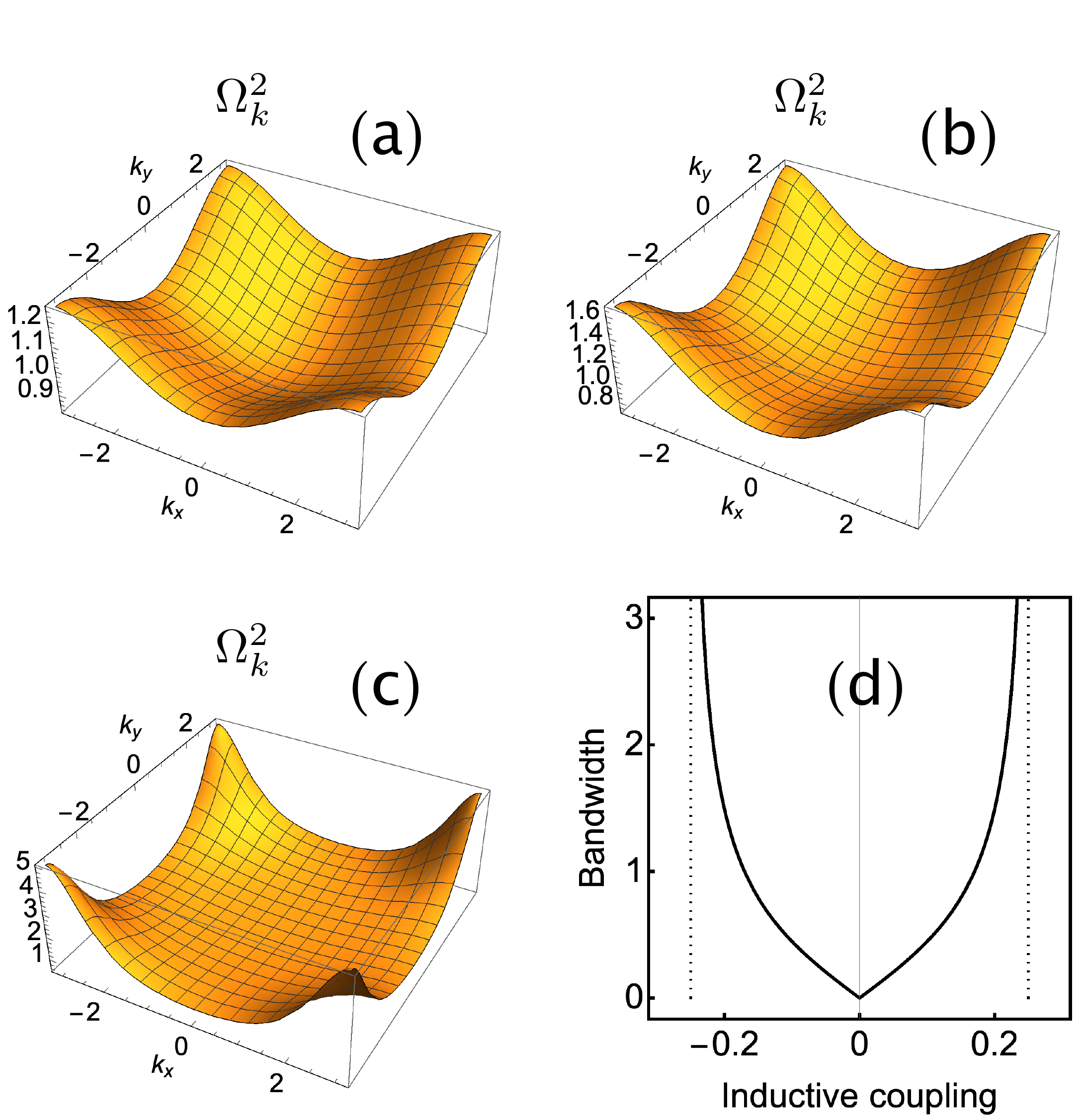}
   \caption{Dispersion relation for homogeneous case: (a) $\lambda=0.05$ (b) $\lambda=0.1$, (c) $\lambda=0.2$ (d) Bandwidth as a function of inductive coupling.}
  \label{fig3}
\end{figure}
Figure~\ref{fig3} shows the band $\Omega_{\bf k}^2$, as well as the bandwidth, defined as 
$|\Omega_{0}^2-\Omega_{\pm\pi}^2|$, is given by $8 \lambda/(1-16 \lambda^2)$.
We see that at the edges of the Brillouin zone, the bandwidth diverges at $|\lambda|=1/4$, but Figs.~\ref{fig3} (a), (b) and (c) show that the increase in bandwidth is mostly concentrated in the immediate vicinity of $(k_{x}, k_{y})=(\pm\pi,\pm\pi)$.

{\em Disorder}.  We introduce now disorder into our system by considering the case when the resonant frequencies $\omega_{\bf n}^2$ are taken as random. This can be done in practice by altering the spacing between the slits, or by inserting different dielectrics in the slits. For the numerical computations of quantities of interest, 
we use a self-expanding square lattice with open boundary conditions. That is, at the
beginning of the computation the lattice is relatively small and grows in size as time evolves, to contain the expanding wavefront. For the evolution times used here, $t\sim 1000$, a typical final lattice dimension is about $64\times 64$, with open boundary conditions. The open geometry is more desirable when one looks to potential applications to magnetic card devices.

Let us get back to the stationary equation (\ref{eq2}) which can be rewritten as
\be
-\left( {1\over{\Omega^2}} \right) q_{\bf n} + \left( {1\over{\omega_{\bf n}^2}} \right) q_{\bf n} + \lambda \left( {1\over{\omega_{\bf n}^2}} \right) \sum_{\bf m} q_{\bf m} = 0
\label{eq13}
\ee
We see right away that the equation corresponds to 
an Anderson tight-binding model, with a site energy term equal to $1/\omega_{\bf n}^2$, and a site-dependent coupling $\lambda/\omega_{\bf n}^2$. That is, the ``diagonal'' term and the ``off-diagonal'' term are completely correlated, and their values appear ``inverted'' when compared to a usual tight-binding model. This implies that the high disorder limit corresponds to small $\omega_{\bf n}^2$, while large $\omega_{\bf n}^2$ values leads to the small disorder limit.

We will explore two kinds of disorder: an uncorrelated one, where the site frequencies $\omega_{\bf n}^2$ are assigned at random from a continuous distribution $[1,\omega^2]$ for a $50-50$ impurity fraction. Clearly, for $\omega^2=1$, we recover the homogeneous case. The second kind of disorder we will explore is a correlated one,  consisting of a generalization to two dimensions of the one-dimensional Random Dimer Model (RDM) of Kundu and Philips\cite{correlated2}. While in the RDM one assigns random site frequencies at pairs (dimers) of lattice sites, in our case we will assign (random) site energies to $4$ nearby sites or `tetramers' of lattice sites. Figure 2 shows an example of a disorder realization for the uncorrelated and correlated cases. We notice the presence of 4-sites clusters that constitute the new, bigger `point impurities'.
\begin{figure}[t]
 \includegraphics[width=1\columnwidth]{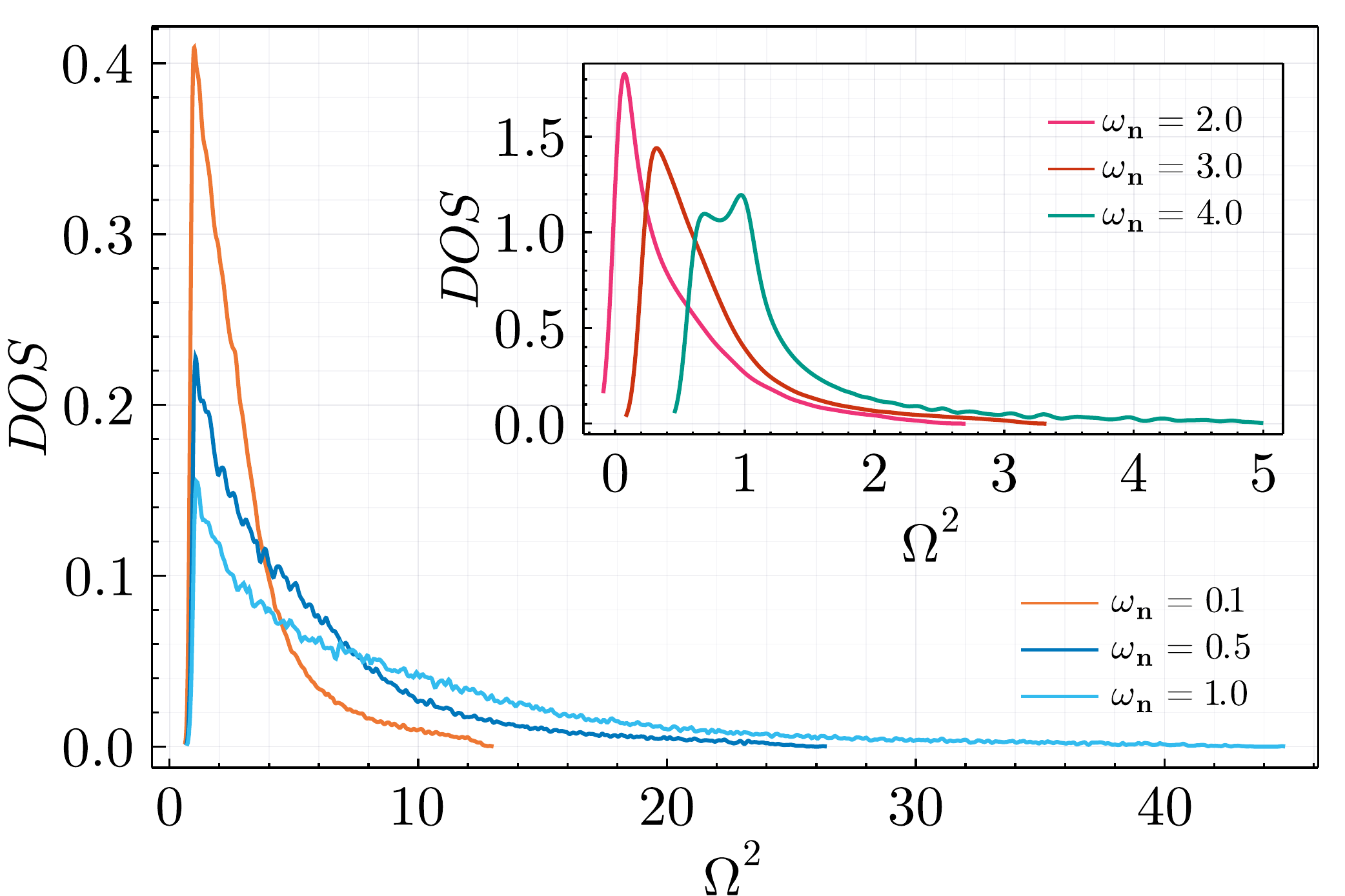}\\
 \includegraphics[width=1\columnwidth]{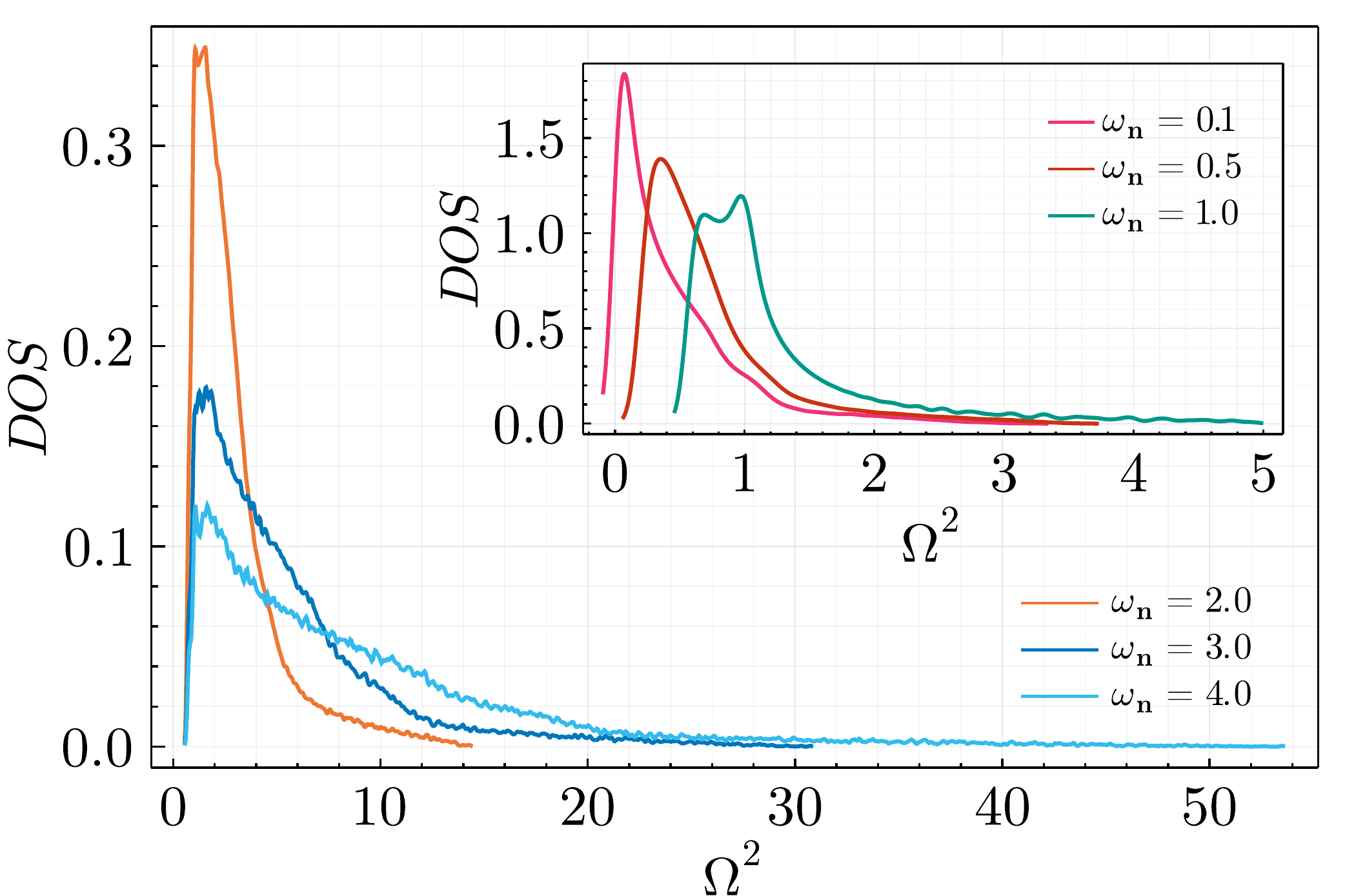}
  \caption{Average density of states for the uncorrelated (top) and correlated (bottom) cases for several impurity width values $\omega$ ($N=30\times 30$, number of realizations=$50$)}
  \label{fig4}
\end{figure}

Figure \ref{fig4} shows the average density of states (DOS)
\be
D(\Omega^2) = \left< (1/N^2)\sum_{m} \delta(\Omega^2-\Omega_{m}^2)\right>
\ee
where $N^2$ is the number of sites, the sum is over all modes, and the average is over all random realizations. The shape of the curve for the ordered case ($\omega^2=1$)  seems to hint at the existence of van Hove-like singularities, which are rounded-off here due to finite-size effects. In both cases, uncorrelated and correlated disorder,  the density of states shows a maximum whose position increases from a small frequency to larger frequencies, as the disorder $w$ is increased. The position of the DOS maxima seems to converge to a fixed frequency value, close to unity. On the other hand, in the limit $\omega\rightarrow 0$, the DOS converges to a Dirac delta function. This can be proven from Eq.(\ref{eq2}) where, for $\omega_{\bf n}^2\rightarrow 0$, the eigenvalue equations reduce to
\be
-\Omega^2 \left( q_{\bf n} + \lambda \sum_{\bf m} q_{\bf m}\right)=0.
\label{xx}
\ee
which implies $\Omega=0$ and $D(\Omega^2)=\delta(\Omega^2)$.

The width of the DOS increases with an increase in disorder, as expected. In general, there is no appreciable difference in the DOS between the disordered monomer and tetramer cases.

\begin{figure}[t]
 \includegraphics[width=1\columnwidth]{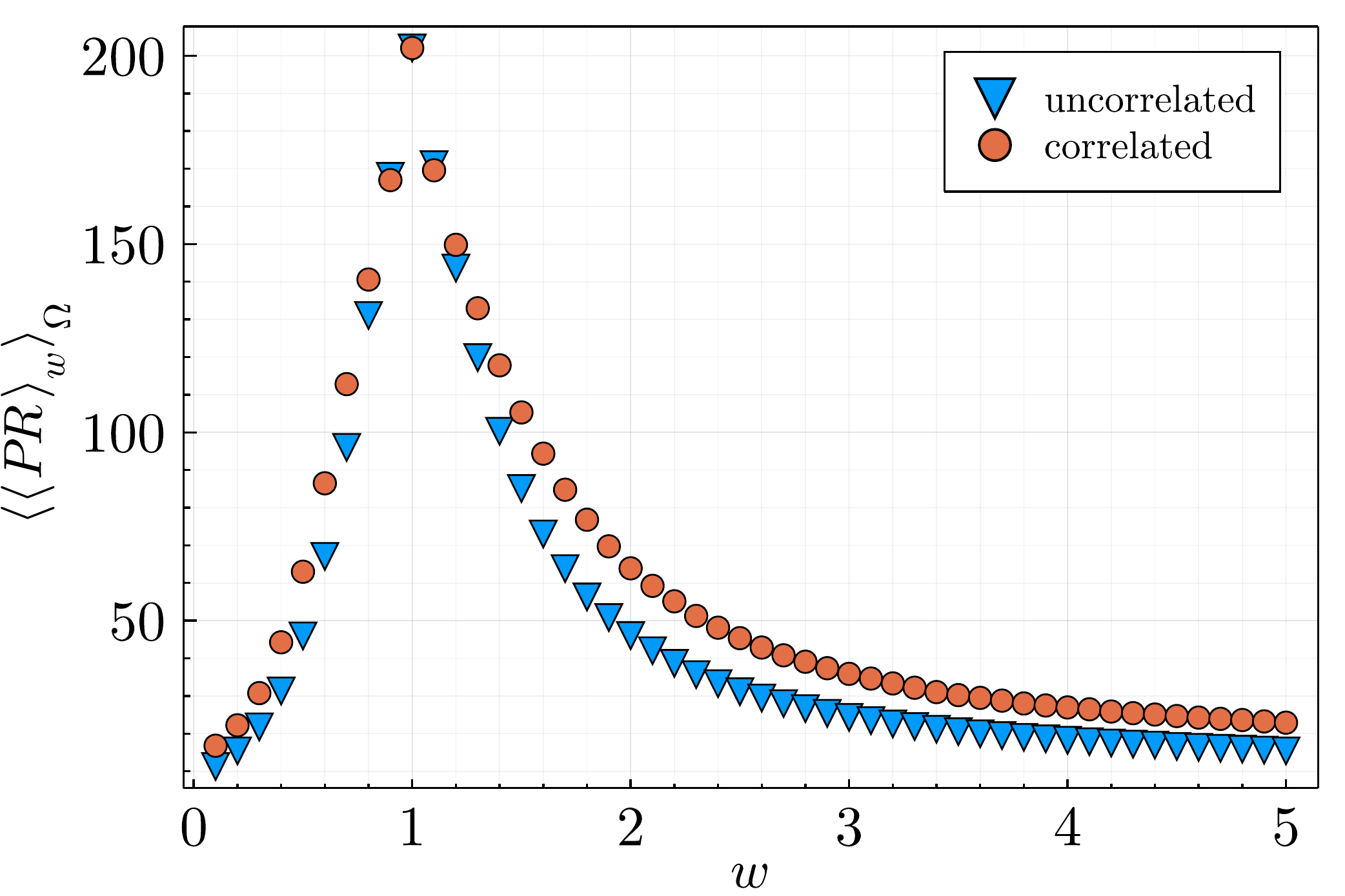}
  \caption{Comparison of the average participation ratio for the uncorrelated (left column) and the correlated case (right column) ($N=22 \times$ 22, number of realizations $= 30.$)}
  \label{fig5}
\end{figure}
For a square lattice of $N\times N$ sites, the participation ratio of mode $\alpha$ is
\be
PR^\alpha=\left( \sum_{n_x, n_y} |q^{\alpha}_{n_x, n_y}|^2\right )^2/\sum_{n_x, n_y} |q^{\alpha}_{n_x, n_y}|^4.
\ee
For a homogeneous  square lattice, we have $q_{\bf n}^{\bf k} \sim \sin(k_{x} n_{x}) \sin(k_{y} n_{y})$, and $PR^{(\alpha)}=(4/9) N \times N$ in the limit of a large number of sites.  Figure 5 shows the average participation ratio $\langle\langle PR \rangle\rangle $ averaged over all mode frequencies and over all disorder realizations,  as a function of disorder width, for the uncorrelated  and correlated cases.  
The overall shape of the $\langle\langle PR \rangle\rangle $ looks quite similar in both cases, with an $\langle\langle PR \rangle\rangle $ that decreases steadily towards both sides around $\omega=1$, the homogeneous case. This is to be expected since, as we commented before, a deviation from $\omega=1$ to smaller $\omega$ values leads to an effective increase in disorder, while something similar happens at values of $\omega$ greater than unity, although in this case the disorder is less drastic than for the other case. Now, we see that the amplitudes of the curves are different for both cases, with the correlated case showing higher $\langle\langle PR \rangle\rangle $ values for all the frequency range. This implies a smaller wave localization, on average than for the uncorrelated case.
\begin{figure}[t!]
 \includegraphics[width = \columnwidth]{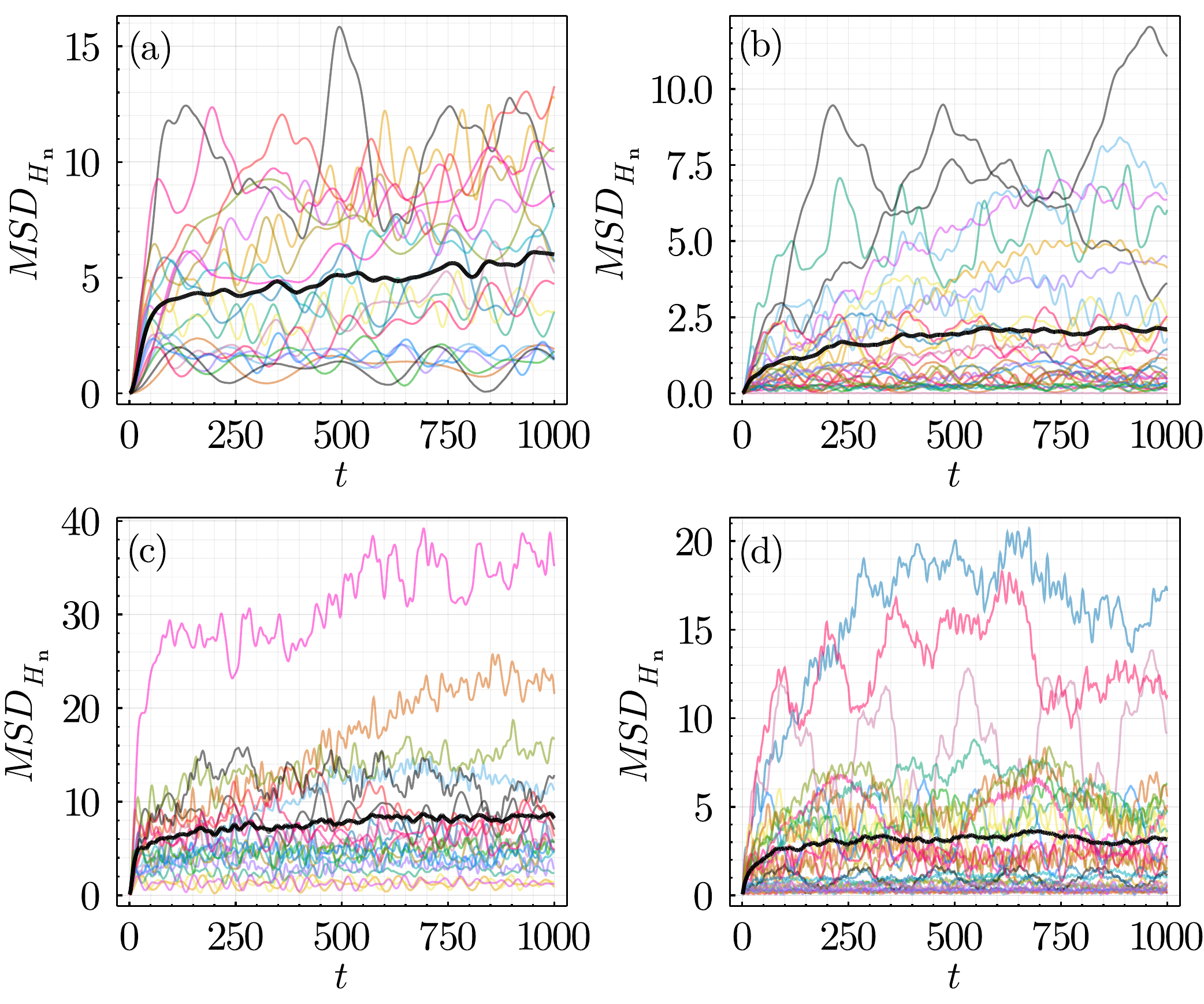}
  \caption{Comparison of the average MSD for the correlated (left column) and the uncorrelated (right column) disorder cases , for several disorder width values. The black curve denotes the average value. (a) and (b): $\omega_{n}\in [0.1, 1]$. (c) and (d): $\omega_{n}\in [1, 4]$
  ($N=64\times 64$, number of realizations=$20$.)}
  \label{fig6}
\end{figure}

\begin{figure}[t!]
 \includegraphics[width = \columnwidth]{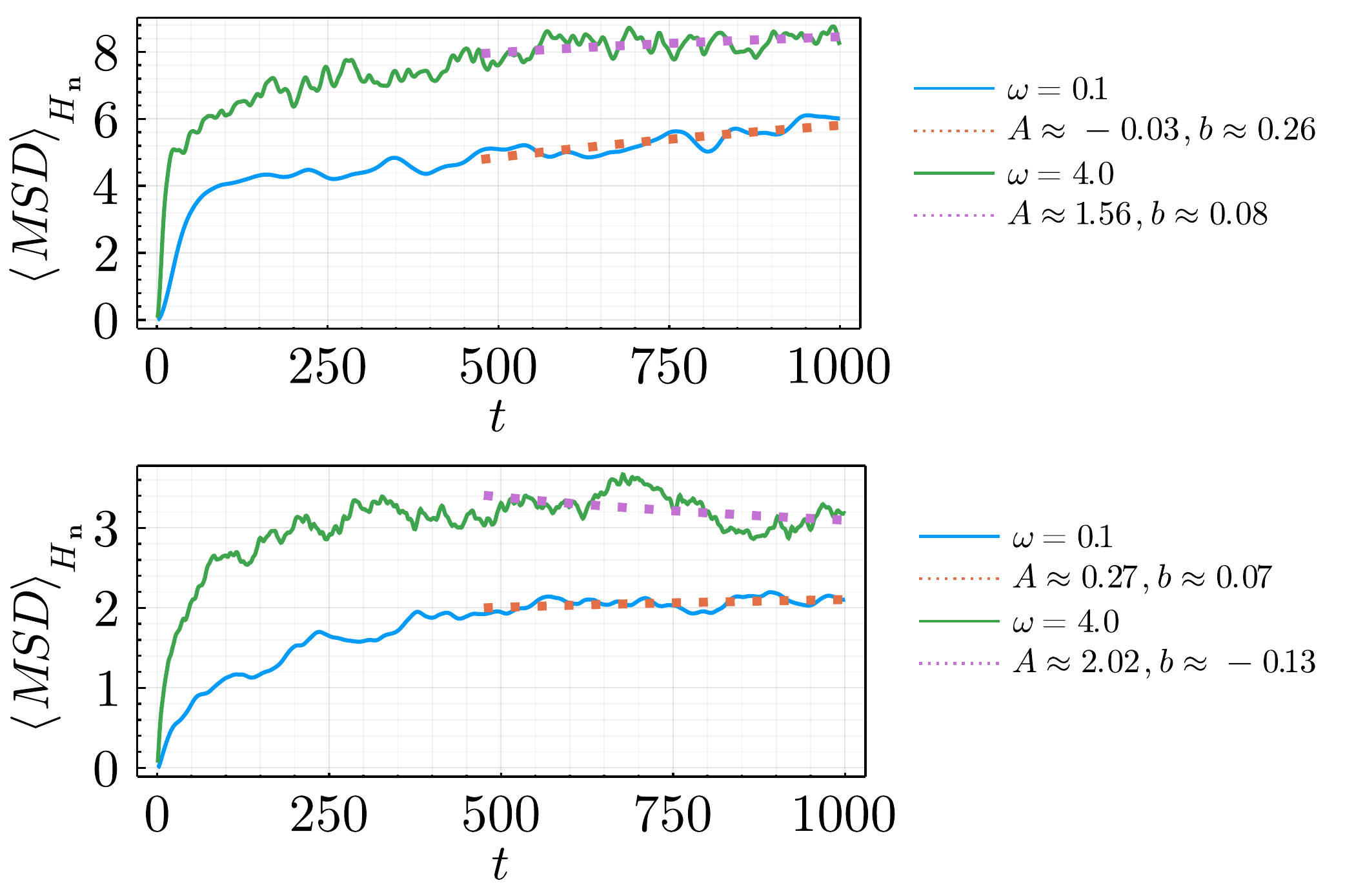}
  \caption{Asymptotic fit $MSD \sim A t^b$ of the average MSD for the correlated 
  (top) and uncorrelated (bottom) cases.}
  \label{fig7}
\end{figure}

{\em Disordered transport}. Let us now compute the spreading of an initially localized magnetic excitation and observe its time evolution. For our SRR array, we will focus
on the the mean square displacement (MSD) of the magnetic energy density
\be
\langle {\bf n}^2 \rangle = \sum_{\bf n} {\bf n}^2 |H_{\bf n}(t)|^2 / \sum_{\bf n} |H_{\bf n}(t)|^2\label{MSD2}
\ee
where $H_{\bf n}$ is given by Eq.(\ref{Hn}). We computed and compared the MSD for the uncorrelated and correlated cases.  Results are shown in Fig.6. We note the presence of strong fluctuations, where the MSD changes considerably from realization to realization. For the uncorrelated  case, the MSD seems to saturate at long times for all disorder widths. A comparison of the MSD for different disorder widths reveals that the saturation value MSD
is greater for the $w>1$ case than for the case with $w<1$. This can be understood from Eq.(\ref{eq13}) where we see that values of $\omega_{\bf n}^2>1$ results in effective low disorder, while values of $\omega_{\bf n}^2<1$ are equivalent to high disorder.
For the tetramers case, we see evidence of a departure from saturation for $\omega<1$, with a sub-diffusive transport $MSD \sim t^{0.26}$ (Fig.7). For $\omega>1$, the MSD 
is greater than its uncorrelated counterpart and also seems to saturate at long times.

Now, what is the origin of the propagating regime we see  for the correlated case? Here, it might be useful to remember that for the one-dimensional traditional Anderson random dimer model (not SRR), there is a nontrivial difference between the uncorrelated and correlated case: The existence of resonant trajectories where a plane wave can propagate across the whole lattice without scattering. This happens for energies close enough to the resonant energy where a wave goes through a dimer with unity transmission. This gives rise to $\sqrt{N}$ of transmitting states\cite{correlated1,correlated2,correlated3} and to a MSD that behaves as $t^{3/2}$ at long times. This effect was also observed numerically in a one-dimensional disordered SRR array\cite{jmmm}. For our 2D case, we could assume for simplicity that a 2D plane wave can be factored as $2$ independent waves that propagate along the horizontal and vertical axes. When this wave encounters a tetramer, it could pass unreflected if
a resonance  condition is obeyed, which involves the contrast between the background frequency ($\omega^2=1$) and the frequency of the tetramer. Since the frequency of the tetramers is extracted from a continuous distribution rather than a binary one, the effect will be small since after traversing a tetramer, the wave might not encounter a favorable transmitting  condition until farther out. Anyway, this small resonance effect decreases the effective disorder of the system, leading to a number of states with larger localization lengths.  This explains the higher $\langle\langle PR \rangle\rangle $ found for the correlated system (Fig.5).

{\em Conclusions}. We have compared the effect of two types of disorder on the eigenmodes and the transport of excitations, of a simplified model of a magnetic metamaterial consisting of a square array of split-ring resonators. The disorder consisted of random values of the resonant frequencies of the rings, taken from a continuous distribution, that are assigned to array sites completely at random (uncorrelated case), or by assigning `tetramers' at random to the array sites (correlated case). The last case corresponds to a generalization of the random `dimer' model for one-dimensional systems of Dunlap and Wu\cite{correlated2}. Our system can be mapped to a tight binding system with site energies and couplings that are nearly identical and with effective disorder strength values that are  `inverted', Eq.(\ref{eq13}). Thus, small (large) values of $\omega_{\bf n}^2$ lead to effective large (small) disorder. The computation of the mean square displacement (MSD) shows that, in general, there is saturation at large times in both cases, but for the correlated case there appears to be a disorder window inside which the MSD increases in time in a sub-diffusive manner. We explained this feature as a possible remnant effect of plane wave resonance across a single tetramer. This conjecture also explains the features of the participation ratio (PR) where for the correlated case we observe the existence of modes whose localization length is greater than for the uncorrelated case.
These results could be of use for the future design of efficient magnetic energy confinement devices, and the harvesting and transport of magnetic energy.

\acknowledgments

This work was supported by Fondecyt Grant 1200120 and the
super-computing infrastructure of the NLHP (ECM-02). We are grateful
to the anonymous referee for his/her valuable suggestions that helped build
a substantially improved paper.

\end{document}